\documentclass[12pt]{article}
\usepackage{amsmath,amsthm,amssymb,multirow,epsfig,subfigure,hhline,rotating}

\def\f{\frac} 
\def\l{\left} 
\def\r{\right}

\begin{document} 
 
\title{\Large \bf Molecular multiresolution surfaces} 

\author{ 
G. W. Wei$^{1,2,}$\footnote{Corresponding author.}, 
Yuhui Sun$^1$, Y. C. Zhou$^1$  and 
M. Feig$^{3}$  
\\ 
$^1$Department of Mathematics \\  
Michigan State University, MI 48824, USA\\
$^2$Department of Electrical and Computer Engineering \\ 
Michigan State University, MI 48824, USA\\
$^3$Department of Biochemistry and 
    Molecular Biology\\
Michigan State University, MI 48824, USA
} 
\date{\today} 
\maketitle 
\begin{abstract} 
The surface of a molecule determines much of its chemical and physical property, and is of 
great interest and importance. In this Letter, we introduce the concept of molecular 
multiresolution surfaces as a new paradigm of multiscale biological modeling. Molecular 
multiresolution surfaces contain not only a family of molecular surfaces, corresponding 
to different probe radii, but also the solvent accessible surface and van der Waals surface as 
limiting cases. All the proposed surfaces are generated by a novel approach, the diffusion map 
of continuum solvent over the van der Waals surface of a molecule. A new local spectral 
evolution kernel is introduced for the numerical integration of the diffusion equation in a 
single time step.

\end{abstract} 

Molecular surface \cite{Richards} is one of the most important concepts in modern 
biochemistry and molecular biology. Albeit molecular surfaces are merely a conventional 
imitation of molecular boundaries, it has been shown that such models can often 
explain fundamental physical and chemical effects. For example, the stability and 
solubility of macromolecules, such as protein, DNA and RNA, are determined by how their 
surfaces interact with solvent and/or other surrounding molecules. Therefore, the structure 
and function of macromolecules depend on the features of their molecular surfaces. 
Important applications of molecular surfaces \cite{Richards}, 
as well as solvent accessible surfaces \cite{Lee}  and van der Waals surfaces, are flaring in 
protein folding \cite{Livingstone,Spolar},  protein-protein interfaces \cite{Crowley}, oral drug 
classification \cite{Bergstrom}, DNA binding and bending \cite{Dragan}, parameterization of heat 
capacity changes \cite{Livingstone}, macromolecular docking \cite{Jackson}, enzyme catalysis 
\cite{LiCata}, calculation of solvation energies \cite{Raschke}, molecular dynamics \cite{Das},
and implicit solvent models \cite{Baker,LeeMS,Rocchia}.

\begin{figure}  
\begin{center}  
\includegraphics[width=0.5\textwidth]{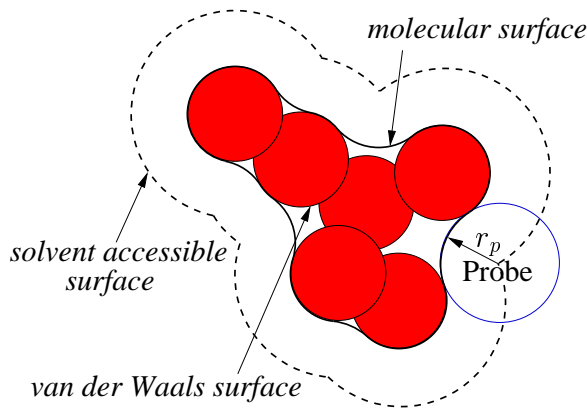} 
\end{center}   
\caption{Illustration of various surfaces.}  
\label{surface}              
\end{figure}

Molecular surface \cite{Richards}, also called solvent excluded surface, is the surface 
that encloses the solvent excluded volume \cite{Connolly}, and can be defined as 
a smooth envelope traced out by the surface of a probe sphere rolled over the molecular 
van der Waals surface.  The latter is a collection of all the unburied 
atomic surfaces of the molecule, see Fig. \ref{surface}. The estimation and analysis of 
solvent accessible surfaces and molecular surfaces have attracted much attention in the 
past three decades after the first calculation of molecular surfaces by Greer and Bush 
\cite{Greer}. A variety of methods have been proposed, including the  Gauss-Bonnet theorem 
\cite{Connolly,Richmond}, closed-form analytical method \cite{Gibson}, space transformation
 \cite{Fraczkiewicz}, alpha shape theory \cite{Liang}, triangulation \cite{Connolly85,Zauhar},
Cartesian grid based methods \cite{Rocchia}, contour-buildup algorithm \cite{Totrov},
binary tree \cite{Sanner}, and parallel methods \cite{Varshney}. Despit of much success, 
in general, calculation of molecular surfaces is both topologically and computationally 
challenging due to the multiple overlap of atomic spheres, overall irregular structure
 of a macromolecule, and self-intersecting singularities \cite{Gogonea,Sanner}. 
The situation where the probe sphere is simultaneously tangent to four atoms poses another 
challenge \cite{Eisenhaber}. Molecular surface calculations are the bottleneck in 
the molecular dynamics of macromolecules, where the molecular surfaces are computed 
billions of times in the course of the simulation.

About four decades ago, Benoi Mandelbrot  posed the profound question of how one measures 
the length of a coastline in his paper entitled ``How long is the coastline of Great Britain-
Statistical self-similarity and fractional dimension'' \cite{Mandelbrot}. The answer 
depends on  what scale the measurement is being made --- one arrives at a different 
length by using a different scale. It is easy to envision that the molecular surface of
a molecule depends the scale of the probe. As biological phenomena occur over a wide 
range of length scales, molecular multiresolution surfaces ought to provide the 
scientific community with a multiscale framework to reveal complex biological processes 
across scales. Unfortunately, such a multiscale framework does not yet exist.

\begin{figure}  
\begin{center}  
\begin{tabular}{cc}   
\includegraphics[width=0.482\textwidth]{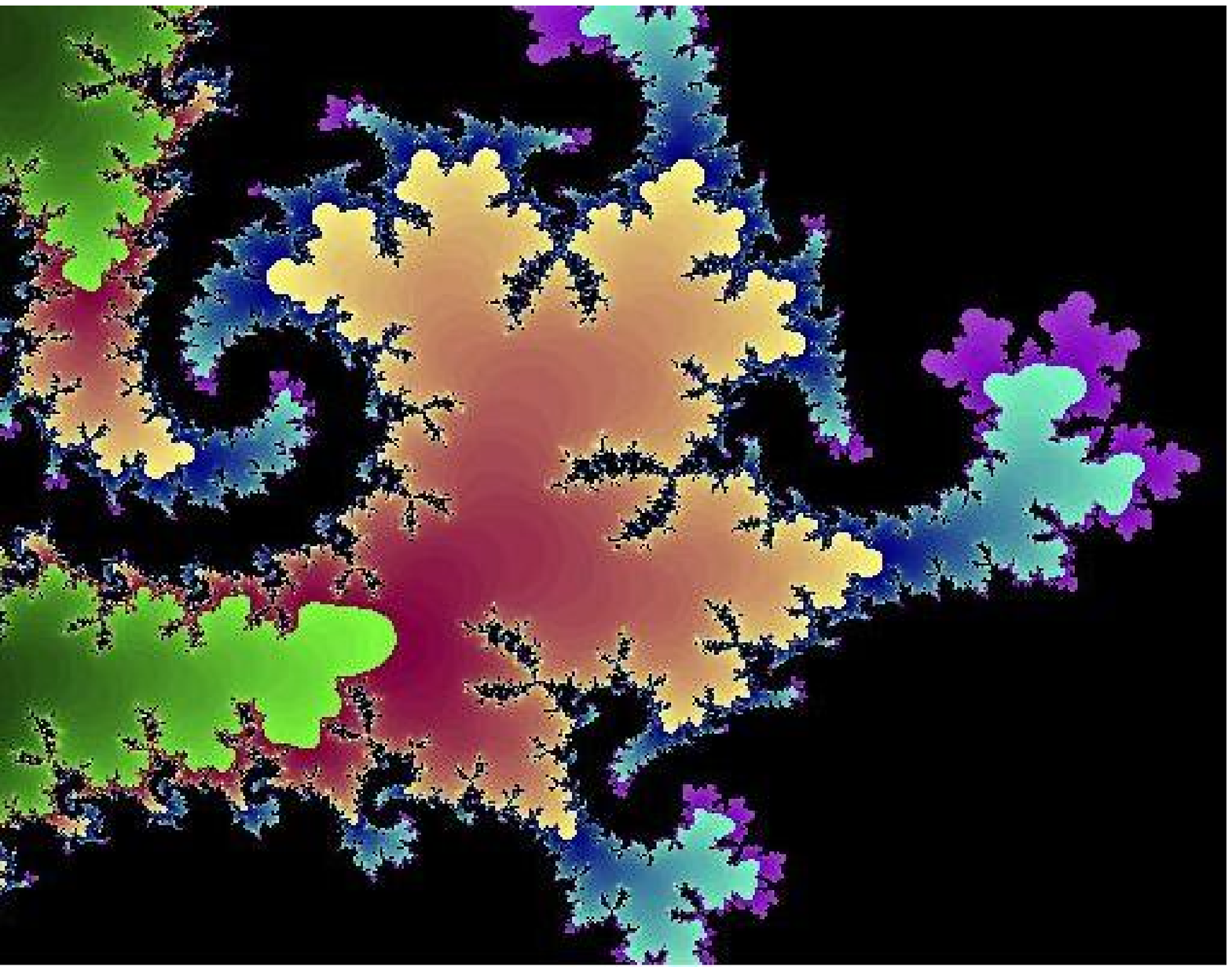} \quad  & \quad  
\includegraphics[width=0.45\textwidth]{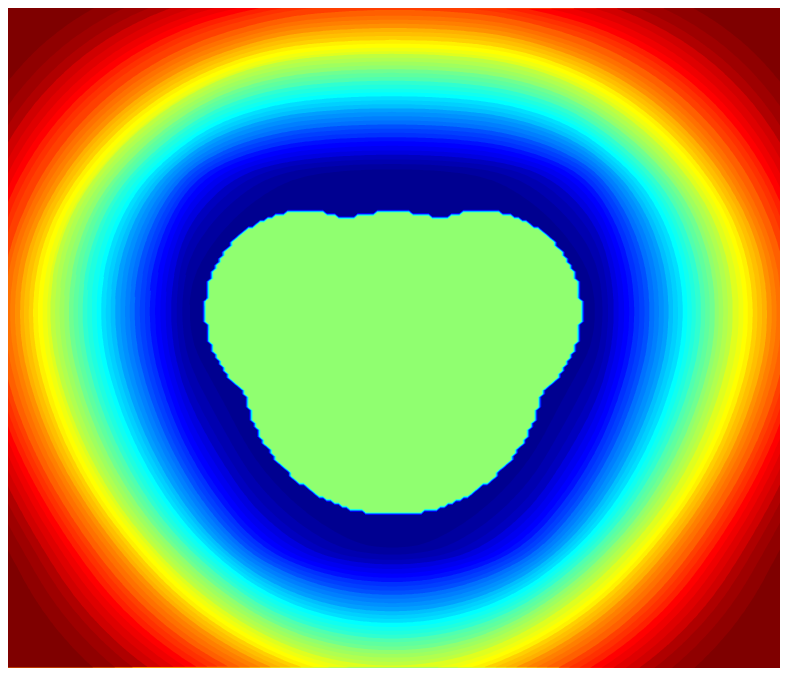}\\ 
(a) & (b) \\
\end{tabular}   
\end{center}   
\caption{
Multiscale surfaces marked up by flow contours.
(a) The diffusion map of a fractal (Courtesy:
     The Mandelbrot Explorer Gallery);
(b) The multiresolution surfaces of a water molecule. 
}  
\label{fractal}              
\end{figure}

The objective of the present work is to introduce the concept of molecular multiresolution 
surfaces both as a multiscale framework for biological modeling and as  a unified 
description of the van der Waals surface, solvent accessible surface and molecular surface. 
The essential idea is similar to the multiresolution profiles of a fractal, generated by
restricted diffusion, see Fig. \ref{fractal}. Use is made to continuum solvent and its 
curvature-controlled diffusion. The mathematical description of the solvent diffusion process 
is formulated by appropriate reaction-diffusion equations. All geometric features of a molecular 
surface, such as convex spherical patches, saddle-shaped pieces of tori, and concave reentrant 
faces, are naturally accounted by the geometric property of partial differential equations with 
appropriate initial-boundary conditions.  A family of molecular multiresolution surfaces are found 
to correspond to different probe radii, and are generated in the continuous distribution of solvent 
density. The van der Waals surface is a special case in which the probe radius vanishes. A fast 
computational algorithm is introduced to generate molecular multiresolution surfaces in a single 
step of time evolution.   

Without the loss of generality, we assume that the fluid flow of continuum solvent has a density 
distribution $\rho({\bf r},t)$ at a space-time location $({\bf r}, t)$, where ${\bf r}\in \mathbb R^3$ 
and $t\in [0,\infty)$.  For an arbitrarily small volume $\Omega$, the motion of the solvent density 
is characterized by density flux ${\bf J}({\bf r},t)$ through the surface of the volume, 
$\partial\Omega$. In this work, we assume that the flux is given by generalized Fick's law  
\begin{equation}\label{heat2}
{\bf J}({\bf r},t)=-\sum_q D_q(K(\rho),{\bf r},t)\nabla \nabla^{2q}\rho({\bf r},t),
\end{equation}
where $D_q(K(\rho),{\bf r},t)$ are diffusion coefficients depending in the Gaussian curvature $K$
of solvent density isosurfaces, and high order terms (i.e., $q>1$) describe the 
influence of inhomogeneity in the solvent density and flux-flux correlation in the density 
flux. In particular, the term $\nabla \nabla^2$ can be regarded as an energy flux operator. 
A similar fourth order flux term was employed in the Cahn-Hilliard equation \cite{Cahn} to describe 
the evolution of a conserved concentration field during phase separation. The total solvent in 
the volume is given by $\int_\Omega \rho({\bf r},t)d{\bf r}$.  The rate of change of the solvent  
in the volume, $\frac{d}{dt}\int_\Omega \rho({\bf r},t)d{\bf r}$, is balanced by the solvent out 
flux,  $-\int_{\partial\Omega}{\bf J}{\bf \cdot}{\bf n({\bf r})}dS$, and solvent
production, $\int_\Omega P(K(\rho),\rho,{\bf r},t) d{\bf r}$
\begin{equation}\label{heat3}
 \int_\Omega   \frac{d}{dt}\rho({\bf r},t)d{\bf r}=-\int_{\partial\Omega}{\bf J}{\bf \cdot}
    {\bf n({\bf r})}dS
+   \int_\Omega P(K(\rho),\rho,{\bf r},t) d{\bf r}
\end{equation}
where ${\bf n({\bf r})}$ is the unit outer normal direction at ${\bf r}$. By using the divergence 
theorem, we convert the surface integral into a volume integral and obtain 
\begin{equation}\label{heat4}
\frac{\partial \rho({\bf r},t)}{\partial t}
=\sum_q \nabla  {\bf \cdot} D_q(K(\rho),{\bf r},t)\nabla \nabla^{2q}\rho({\bf r},t)
+   P(K(\rho), \rho, {\bf r},t).
\end{equation}  
The diffusion coefficients $D_q(K(\rho),{\bf r},t)$ are position dependent because of 
the possible presence of inhomogeneity, such multivalent ions \cite{Rocchia01}, and their  
curvature-dependence is designed for the geometric control of solvent density isosurfaces 
in the final diffusion map. For example, the solvent density distribution near the molecular
surface can be made to preserve the curvature of the probe molecule by taking 
$D_q(K(\rho),{\bf r},t)\sim \exp\left(-\frac{(K-K_p)^2}{2\kappa_q^2}\right)$, where $K_p$ is 
the Gaussian curvature of the probe  and $\kappa_q$  are constants. The solvent production 
can be either negative or positive, and is designed for further geometric control of the 
molecular surface. It can also be used to add or remove atoms at specific locations. 
High order gradient-controlled diffusion equations were proposed by Wei \cite{weiieeespl} 
for efficient image processing. Anisotropic reaction-diffusion equations have been introduced 
for shock-capturing in the context of computational fluid dynamics \cite{weicpc2002}.  

In this work, we define the problem of finding molecular multiresolution surfaces as an 
initial-boundary value problem governed by the new curvature-controlled diffusion equation (\ref{heat4}) 
in a domain $\Omega$ that is large enough to include all the surfaces of interest of a molecule. 
The Direchlet boundary condition is employed at domain boundary $\partial \Omega$, i.e., 
$\rho({\bf r},t)=\rho_0,~~~~{\bf r}\in\partial \Omega$. Initially, there is no presence of 
solvent in the  domain, i.e., $\rho({\bf r},0)=0,~~~~ {\bf r}\in\Omega$. 
We set $D_q(K(\rho),{\bf r},t)=0$ indiscriminately in all the atomic volumes 
$V_i=\frac{4}{3}\pi r_{i}^3$, where $r_i$ is the $i^{\rm th}$ atom of the molecule. 
We choose the evolution time sufficiently long so that the solvent density can reach the center 
of the computational domain. The diffusion of the solvent toward to the domain center leads to 
a family of profiles of solvent density distribution, which in turn gives rise to a family of 
molecular multiresolution surfaces, corresponding to different solvent probe radii $r_p$ and 
different resolutions. Note that the van der Waals surface, $\partial \Omega_{VWS}$, corresponds
to $r_p=0$, and is obtained at the level set $\rho=0$.  Here we define the midway molecular 
surface as the multiresolution molecular surface whose convex surfaces locate at the middle 
of the van der Waals surface and the solvent accessible surface. Such a surface should be 
useful as it unclear whether the solvent accessible surface or the solvent excluded surface
better describes  hydration effects \cite{Jackson}. In general, the nonlinear terms 
in Eq. (\ref{heat4}) can be utilized to provide large gradient near the desirable molecular surface. 
This approach directly results in a Cartesian grid representation of the molecular surface. Such 
a representation is important for many applications, particularly for implicit solvent modeling 
of electrostatics in macromolecules \cite{Rocchia}.

\begin{figure}  
\begin{center}  
\begin{tabular}{cc}   
\includegraphics[width=0.35\textwidth]{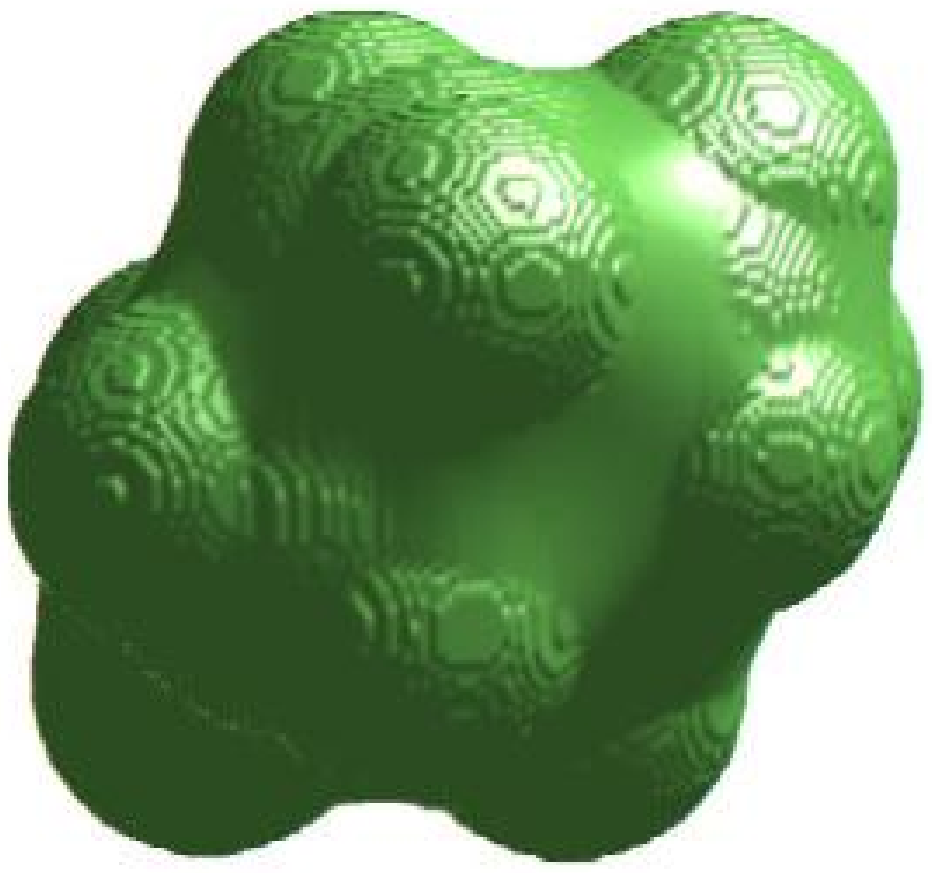} \quad \quad & \quad  \quad  
\includegraphics[width=0.45\textwidth]{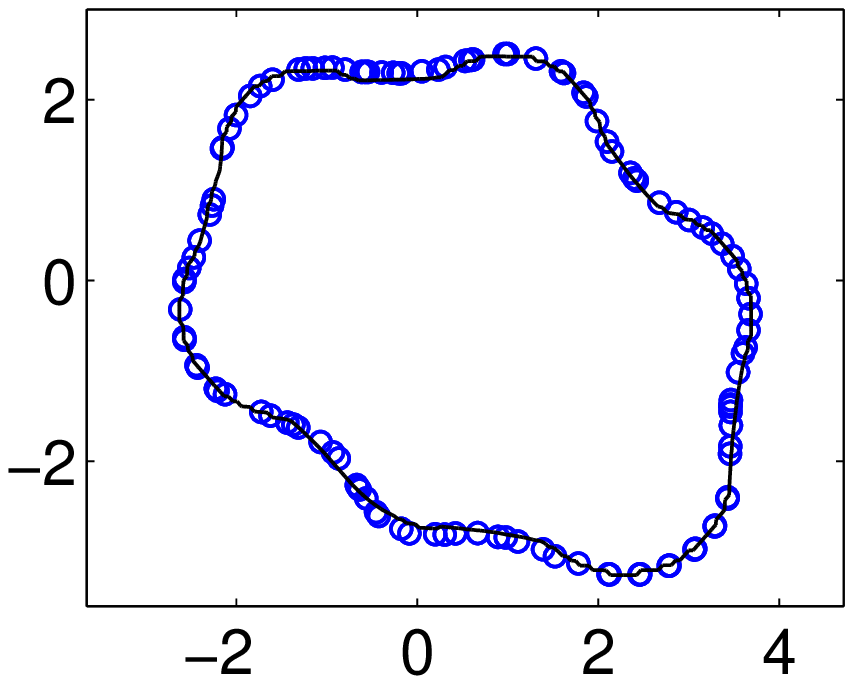}\\ 
(a) & (b) \\
\includegraphics[width=0.45\textwidth]{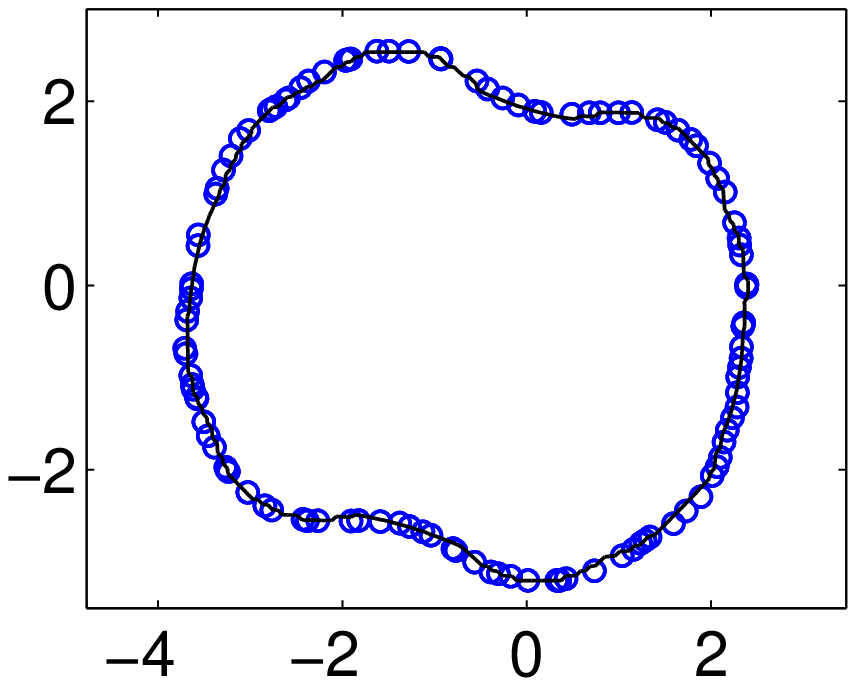} \quad \quad & \quad  \quad  
\includegraphics[width=0.45\textwidth]{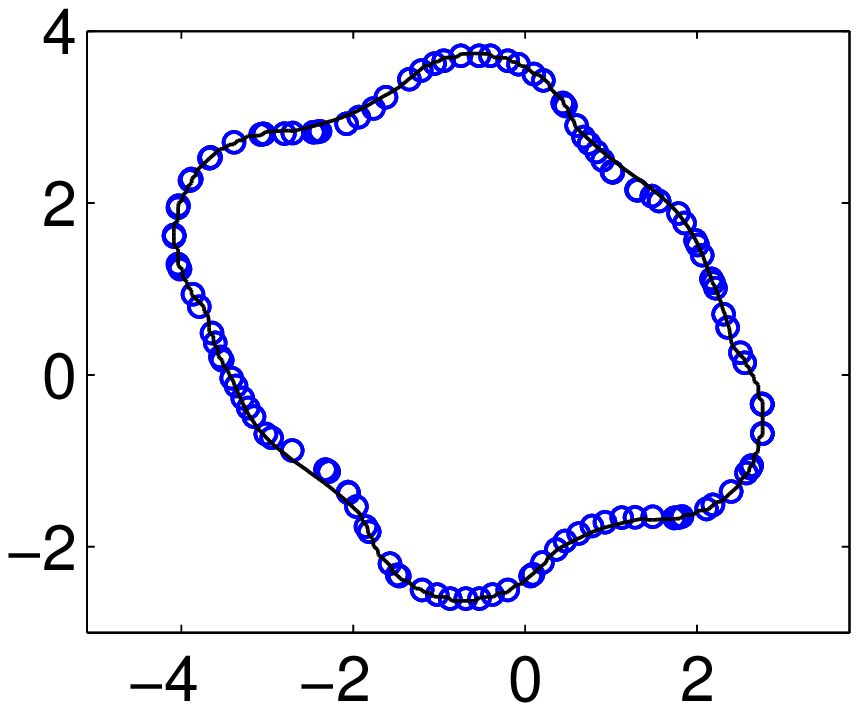}\\ 
(c) & (d) \\
\end{tabular}   
\end{center}   
\caption{Molecular multiresolution surfaces of cyclohexane and their comparison with those
obtained by the MSMS (Solid lines: present; Circles: MSMS).
(a) The solvent excluded surface; 
(b) The cross section at $x=0.6$;
(c) The cross section at $y=0.6$;
(d) The cross section at $z=0.6$.}  
\label{gram}              
\end{figure}

In general, it might not be convenient to solve the nonlinear reaction-diffusion equation 
(\ref{heat4}). Thus simplifications are desirable. For homogeneous solvent, it is sufficient to 
set $D_1$ a piecewise constant $D_1=D({\bf r})$, and $D_q=0$ if  $q\neq 1$. For simplicity, we 
also drop the production term in the rest of discussion. Therefore, we arrive at a much simpler 
problem of diffusion 
\begin{eqnarray}\label{heat5}
&& \frac{\partial \rho({\bf r},t)}{\partial t}=\nabla {\bf \cdot} D({\bf r})\nabla \rho({\bf r},t), 
    ~~~ {\bf r}\in\Omega\\
&& \rho({\bf r},t)=\rho_0,~~~~{\bf r}\in\partial \Omega \nonumber \\ 
&& \rho({\bf r},0)=0, ~~~~{\bf r}\in\Omega \nonumber  \\ 
&& D({\bf r})=1, ~~~~{\bf r}\in\Omega/ V_{VWV}, \nonumber \\
&& D({\bf r})=0, ~~~~{\bf r}\in V_{VWV}, \nonumber
\end{eqnarray}  
where $V_{VWV}$ is the molecular van der Waals volume given by $V_{VWV}=\bigcup_iV_i$. However, the 
resulting diffusion equation does not yet admit a closed-form solution  due to the irregular atomic 
surface information in $D({\bf r})$. Fortunately, the numerical solution of Eq. (\ref{heat5}) is 
quite robust. For example, both implicit and explicit approaches, as well as various spatial 
discretizations can be used to generate the solvent density profiles near the van der Waals 
surface.

The van der Waals surface generated by this approach is exact in the sense that its resolution 
is only limited by the finite mesh size. However, the molecular surface, or the solvent excluded 
surface computed in this manner could be distorted because the different diffusion path 
length for each different part of a molecular. To eliminate this error of molecular anisotropy, 
we set the computational boundary to be the solvent accessible surface ($\partial 
\Omega =\partial \Omega_{SAS}$). To this end, we define solvent-accessible radius of 
$i^{\rm th}$ atom as the sum of its atomic radius and the probe radius, $r_{i}^{SAR}=r_i+r_p$. 
The volume associated with solvent-accessible radius of the $i^{\rm th}$ atom is given
by $V_i^{SAR}=\frac{4}{3}\pi \left[r_i^{SAR}\right]^3$. We initialize the solvent density 
on the desirable solvent accessible surface by  setting $\rho({\bf r},0)=0, ~~~\forall {\bf r}\in 
\bigcup_i V_i^{SAR}$ and $\rho({\bf r},0)=\rho_0$ in rest of the domain. A large probe radius $r_p$ 
can be chosen to ensure a variety of molecular multiresolution surfaces. Note that solvent 
accessible surface marked out by the solvent density $\rho({\bf r},0)$ is also exact.

\begin{figure}  
\begin{center}  
\includegraphics[width=0.4\textwidth]{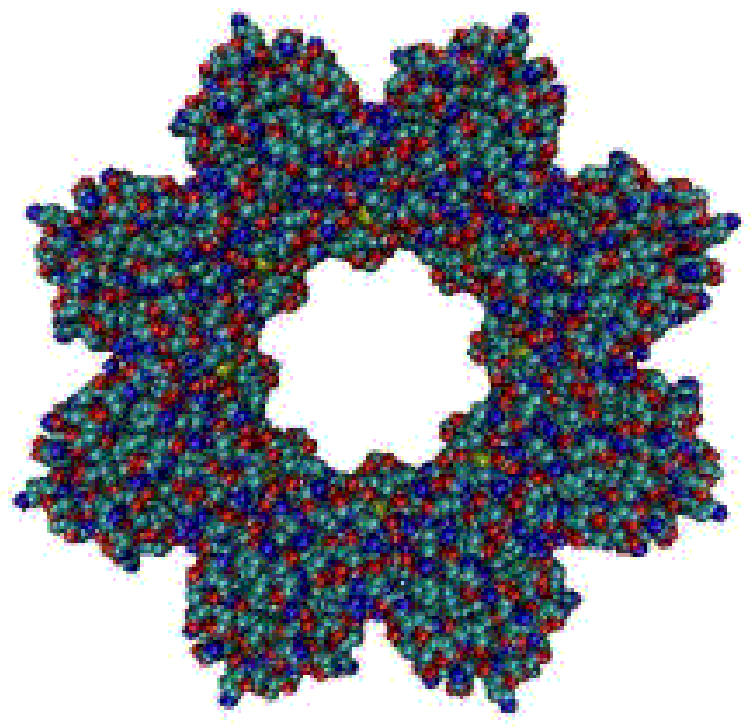} \\
\includegraphics[width=0.4\textwidth]{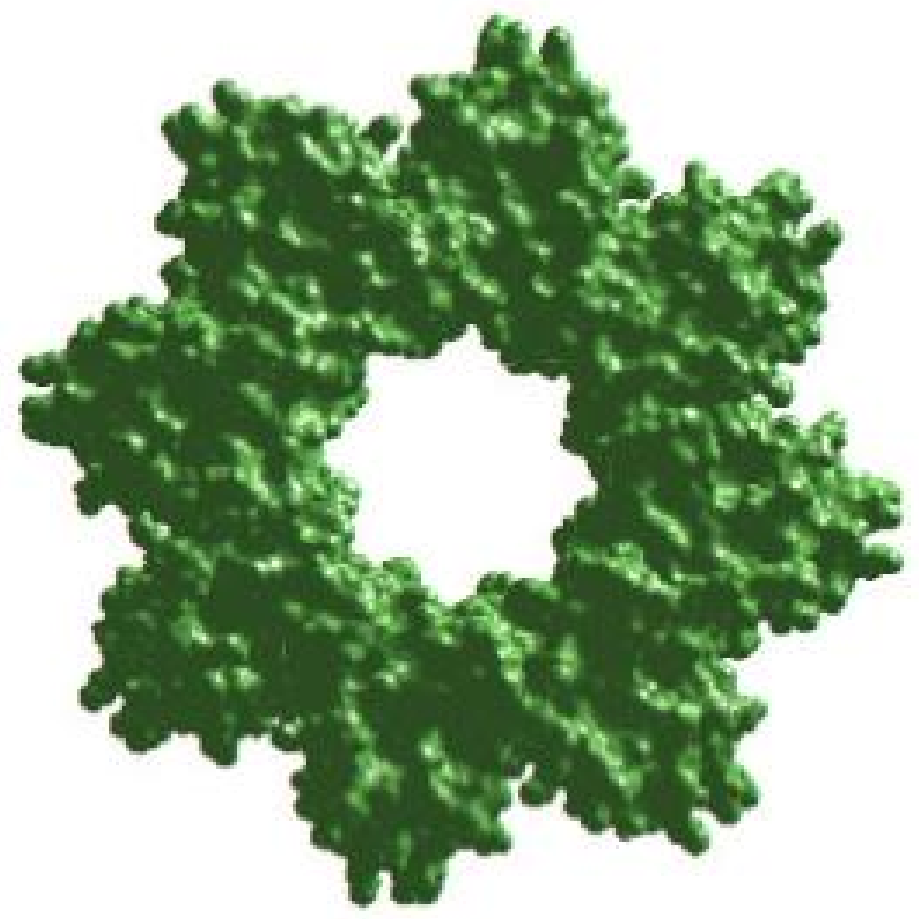}\\ 
\includegraphics[width=0.4\textwidth]{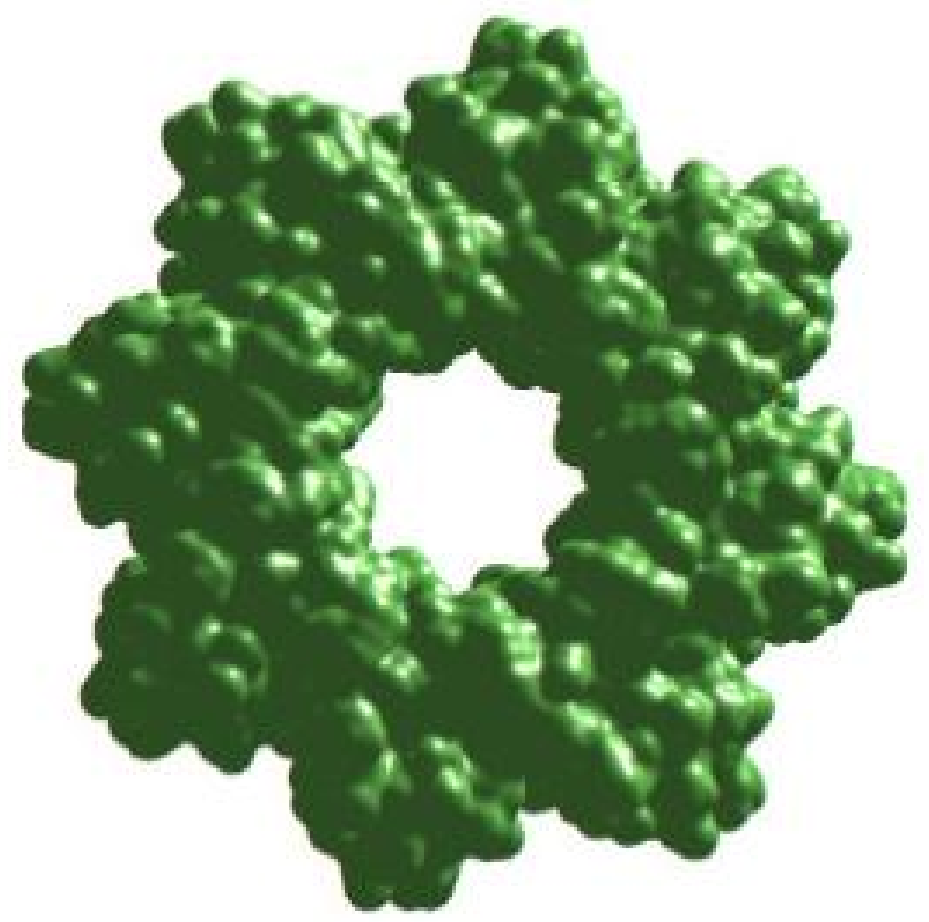} 
\end{center}   
\caption{Molecular multiresolution surfaces of the cell division protein. 
Top: The van der Waals surfaces;
Middle: The solvent excluded surface;
Bottom: The midway molecular surface.}
\label{MJ}              
\end{figure}

To speed up the solution of Eq. (\ref{heat5}), we introduce a fast local spectral evolution 
kernel (LSEK) method \cite{yu2005}, which is able to analytically solve a class of 
reaction-diffusion-convection equations  
$
\frac{\partial}{\partial t}\rho =\left[D(t)\frac{\partial^2} {\partial x^2} 
+ C(t)\frac{\partial} {\partial x} 
+P(t)\right]\rho 
$
in a single time step. The LSEK is exact in time and is of controllable accuracy in space. 
Its complexity  at a fixed level of accuracy scales as $O(N)$, where $N$ is the number of 
grid points. The solution of Eq. (\ref{heat5}) at an arbitrary point of space-time, 
$(x_i,y_j,z_k,t)$, is obtained in a finite difference manner 
\begin{eqnarray}\label{eqapp2}
	&&\rho(x_i,y_j,z_k,t)=\sum_{l_x=-M_x}^{M_x}
	                    \sum_{l_y=-M_y}^{M_y}
	                    \sum_{l_z=-M_z}^{M_z}
	        K_{h_x,\sigma_x}(l_xh_x,t) \nonumber \\ && 
          K_{h_y,\sigma_y}(l_yh_y,t)
          K_{h_z,\sigma_z}(l_zh_z,t)
          \rho(x_i-l_xh_x,
               y_j-l_yh_y,
               z_k-l_zh_y,
               0),
\end{eqnarray}
where $2M_\alpha+1$ are the stencil width for $\alpha=x,y,z$,~ $h_\alpha$  the grid 
spacings, and $K_{h,\sigma}(x,t)$  the LSEK  \cite{yu2005} 
\begin{eqnarray}\label{eqaxtl}
	K_{h,\sigma}(x,t)=\f{h}{\sigma}
\sum_{n=0}^{M_h/2}\l(-\f{1}{4}\r)^n\f{1}{\sqrt{2\pi}n!}
\l(\f{\sigma}{\sigma_{t}}\r)^{2n+1} h_{2n}\l(\f{x}{\sqrt{2}\sigma_{t}}\r).
\end{eqnarray}
Here $h_{2n}(x)$  is the Hermite function defined by the Hermite polynomial $H_{2n}$, 
i.e., $h_{2n}(x)=\exp(-x^2)H_{2n}(x)$, $M_h$ is the highest degree of Hermite 
polynomial used, $\sigma$ is window parameter and $\sigma_{t}^2=\sigma^2+2\int_0^tD(s)ds$.
We set $M_h=88, M_\alpha=32$  and $\sigma_\alpha=3.05h_\alpha$ in this work. 

As a proof of principle, we present two examples. In the first example, cyclohexane (C$_6$H$_{12}$),
a test case recommended by Sanner {\it et al.} \cite{Sanner}, is considered to validate the proposed 
method for calculating the solvent excluded surface. The van der Waals radii of the carbon and 
hydrogen atoms are taken as 1.7 and 1.2 A, respectively. We choose a grid of 200 points 
in each dimension and set $\rho_0=100, r_p=1.5$ A. The solvent excluded surface is extracted 
at $\rho({\bf r},12)=0.04$, see Fig. \ref{gram}a. We also compute the solvent excluded surface 
of  cyclohexane by using the MSMS algorithm \cite{Sanner}. A comparison of three cross sections 
generated from the present approach and those from the MSMS is depicted in Figs. \ref{gram}b-d. 
There is an excellent agreement among these results. 

We next consider a cell division and cell wall biosynthesis protein \cite{Chen}. It has  9772 
atoms in eight polymer chains and 1328 residues. We choose the same set of computational 
parameters as those described in the preceding paragraph. Three multiresolution molecular surfaces, 
the van der Waals surface, the the solvent excluded surface, and the midway molecular 
surface of the cell division protein are depicted in Fig. \ref{MJ}. It is seen that the midway 
molecular surface captures the symmetry and other main features of the protein structure while 
avoids unnecessary details.

In summary, we have introduced a new concept, molecular multiresolution surfaces, 
for the multiscale modeling of biomolecules. The proposed concept provides a unified 
description of the van der Waals surface, solvent accessible surface and solvent excluded 
surface. A novel approach based on the controlled diffusion of continuum solvent 
density is proposed to generate the multiresolution surfaces. A fast local spectral evolution 
kernel is introduced to integrate the diffusion equation in a single time step. The proposed
method does not need to discriminate buried atomic surfaces from unburied ones during 
the computation. It by passes the difficulty of singularities in other existing methods. 
Numerical experiments are carried out to validate and demonstrate the proposed approach.
The formulation and examples given here hopefully will spur the interest of other 
investigators in this and related fields.

\end{document}